\documentclass[a4paper]{jpconf}
\usepackage{graphicx}
\usepackage{float}
\usepackage{epsfig}
\usepackage{graphicx}
\usepackage{array}
\usepackage{float}
\usepackage{rotfloat}
\usepackage{amstext}
\usepackage{graphicx}
\usepackage{rotating}
\usepackage{amssymb}
\usepackage{color}
\usepackage[utf8]{inputenc}

\begin{document}
\title{Micromegas-TPC operation at high pressure in Xenon-trimethylamine mixtures}

\author{D. C. Herrera$^a$, S. Cebrián$^a$, T. Dafni$^a$, E. Ferrer-Ribas$^b$, I.~Giomataris$^b$, D. Gonzalez-Diaz$^{a,c}$,
 H.~Gómez $^{a,d}$, F.J. Iguaz$^a$, I.G.~Irastorza$^a$, G. Luzon$^a$, A. Rodríguez$^a$,
 L.~Segui$^a$, A. Tomás$^a$}
\address{$^a$ Laboratorio de Física Nuclear y Astropartículas, Universidad de Zaragoza, Zaragoza, Spain}
\address{$^b$ IRFU,Centre d’Etudes Nucleaires de Saclay (CEA-Saclay) 91191 Gif sur Yvette, France}
\address{$^c$ Department of Engineering Physics, Tsinghua University, Liuqing building, Beijing, China}
\address{$^d$ Laboratoire de l'Accélérateur Linéaire, Université Paris-Sud 11, Orsay Cedex (FRANCE)}
\ead{diana.he@unizar.es}

\begin{abstract}
We present in this work measurements performed with a small Micromegas-TPC using a xenon-trimethylamine (Xe-TMA) Penning-mixture as filling gas.
Measurements of gas gain and energy resolutions for $22.1$ keV X-rays are presented,
spanning several TMA concentrations and pressures between $1$ and $10$ bar.
Across this pressure range, the best energy resolution and largest increase in gain at constant field (a standard figure for characterizing Penning-like energy transfers)
is observed to be in the $1.5\%$-$2.5\%$ TMA region.
A gain increase (at constant field) up to a factor $100$ and a best energy resolution improved by up to a factor 3 with respect to the one  
previously reported in pure Xe -operated Micromegas, can be obtained. In virtue of the VUV-quenching properties of the mixture, the overall 
maximum gain achievable is also notably increased (up to 400 at 10bar), a factor $\times 3$ higher than in pure Xe. 
In addition, preliminary measurements of the electron drift velocity in a modified setup have been performed and show good agreement with the one obtained from Magboltz.

These results are of great interest for calorimetric applications in gas Xe TPCs,
in particular for the search of the neutrino-less double beta decay (0${\nu\beta\beta}$) of $^{136}$Xe.
\end{abstract}

\section{Introduction}

The detection of the 0${\nu\beta\beta}$ decay in $^{136}$Xe can provide
both the neutrino mass scale and an unambiguous answer to its nature (Majorana or Dirac).
For this reason, the development of large gas Xe Time Projection Chambers (TPCs)
targeted at finding evidence for 0${\nu\beta\beta}$ in $^{136}$Xe is currently active.
This kind of detector can fulfill the main requirements of the current generation of experiments (100~kg~Xe),
which are an energy resolution down to $1\%$ FWHM (full width half maximum) at the excess energy of the decay
(Q$_{\beta\beta}$), and topological information from track reconstruction.
Moreover, it can be easily scaled up to 1 ton, thus allowing for the exploration of the entire inverse hierarchy of neutrino mass models.
The NEXT experiment (Neutrino Experiment with Xenon TPC) will search the 0${\nu\beta\beta}$ decay of ${^{136}}$Xe at the Canfranc Underground Laboratory using
100 kg of enriched Xe in an electroluminescent high pressure TPC~\cite{NEXT,CarlosNEXT}. As part of the R\&D program of the NEXT experiment,
various prototypes were built, one of them based on charge collection from ionization using Micromegas detectors \cite{NEXT-TPC2010,NEXT-Pisa}.
Specifically, the microbulk technique \cite{MicroTech} has been used in several experiments due to its excellent position resolution and good energy resolution, as well as
robustness and low radioactivity \cite{Paco,Radiopurity}, being this latter feature more important for rare events searches \cite{T-REX}.

We focused on the possibility of improving the energy resolution of the Micromegas-TPC using
Xe-TMA Penning mixtures. Early studies performed in wire chambers at 1 bar \cite{XeBas} showed
indeed a great gain enhancement and improved energy resolution.
We performed an experimental study of this mixture which, it is believed, can also provide a starting point for evaluating the recent proposal of using TMA as additive gas to reduce the Fano factor in Xe-based TPCs \cite{Nygren}.
All results conveyed in this work have been obtained with a small-size TPC, that is later described. A medium-size $80l$-TPC (NEXT-MM) has been built and commissioned based on the microbulk technology, and its preliminary tracking performances presented at the conference. This topic will be the subject of a dedicated paper.

The main characteristics of the experimental setup are described in section \ref{section:ExSetUp},
while in section \ref{section:NEXT-0} and \ref{section:NEXT-0-MMsilicon} the experimental procedure and measurements are summarized.
Finally, in section \ref{section:Conclusions} the conclusions and outlook is presented. A more detailed description of these measurements can be found in \cite{Diana}.

\begin{figure}[hb]
\begin{center}
\includegraphics[scale=0.35]{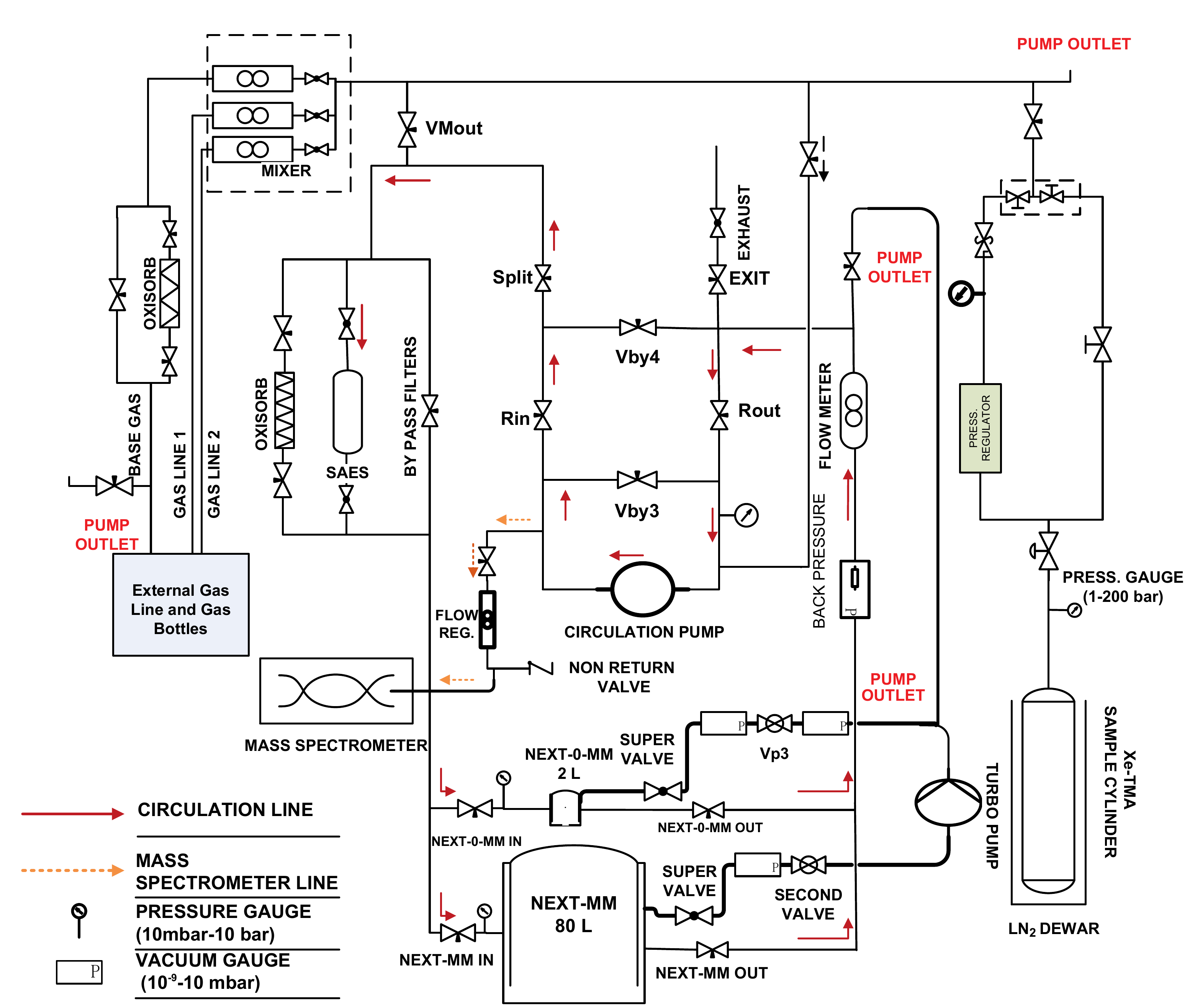}
\end{center}
\caption{Schematic view of the experimental setup, underlining the gas system.  At the bottom/center, the TPC employed for these measurements (dubbed `NEXT-0') is shown. A medium-size TPC (NEXT-MM), central to this system, will be subject of a forthcoming communication.
\label{fig:ExperimentalSetup}}
\end{figure}

\section{Experimental setup}\label{section:ExSetUp}

The experimental setup consists of a small TPC and a mass spectrometer -monitored gas system that allows, amongst other features, for high-pressure recirculation, gas filtering and cryo-pumping.
A schematic view of the experimental setup, mainly underlining the gas system, is shown in Fig. \ref{fig:ExperimentalSetup}, where the small TPC has been dubbed `NEXT-0', described in more detail in \cite{Paco}.
The TPCs inserted in the gas system are built of stainless steel with materials of low outgassing and have been tested up to 12 bar.
The small TPC has an inner volume of $2.4$ l ($10$ cm height, $16$ cm diameter).
The drift distance, equipped with copper rings, can be varied between $1$ and $6$ cm.

For these measurements, a special purifier (SAES 702) was installed in order to work with TMA,
therefore water vapor and electronegative impurities (H$_{2}$O, O$_{2}$, CO$_2$) were constantly removed during the recirculation process.
The gas recovery system has a stainless steel sample cylinder of $2.2$ l
that is immersed in a Dewar flask filled with liquid nitrogen (LN$_2$) during the recovery process.
Finally, a Pfeiffer OmniStar mass spectrometer is used to quantify the gas composition of
Xe+TMA mixtures and to monitor the electronegative impurities.
\section{Experimental procedure and results}\label{section:NEXT-0}

The first experimental goal was to establish a range of TMA concentration for which it would
be possible to obtain the best energy resolution and the highest gain for pressures between $1$ and $10\,\mbox{bar}$.
Hence, a systematic variation of the TMA concentration was performed at four reference pressures: $1$,
$5$, $8$ and $10\,\mbox{bar}$ (see section \ref{subsection:varTMA}).
Once the optimal TMA concentration range was established, a systematic study at various pressures
in a range from $1$ to $10$ bar was realized (see section \ref{subsection:VarPre}).

\begin{figure}[ht]
\begin{center}
\includegraphics[scale=0.35]{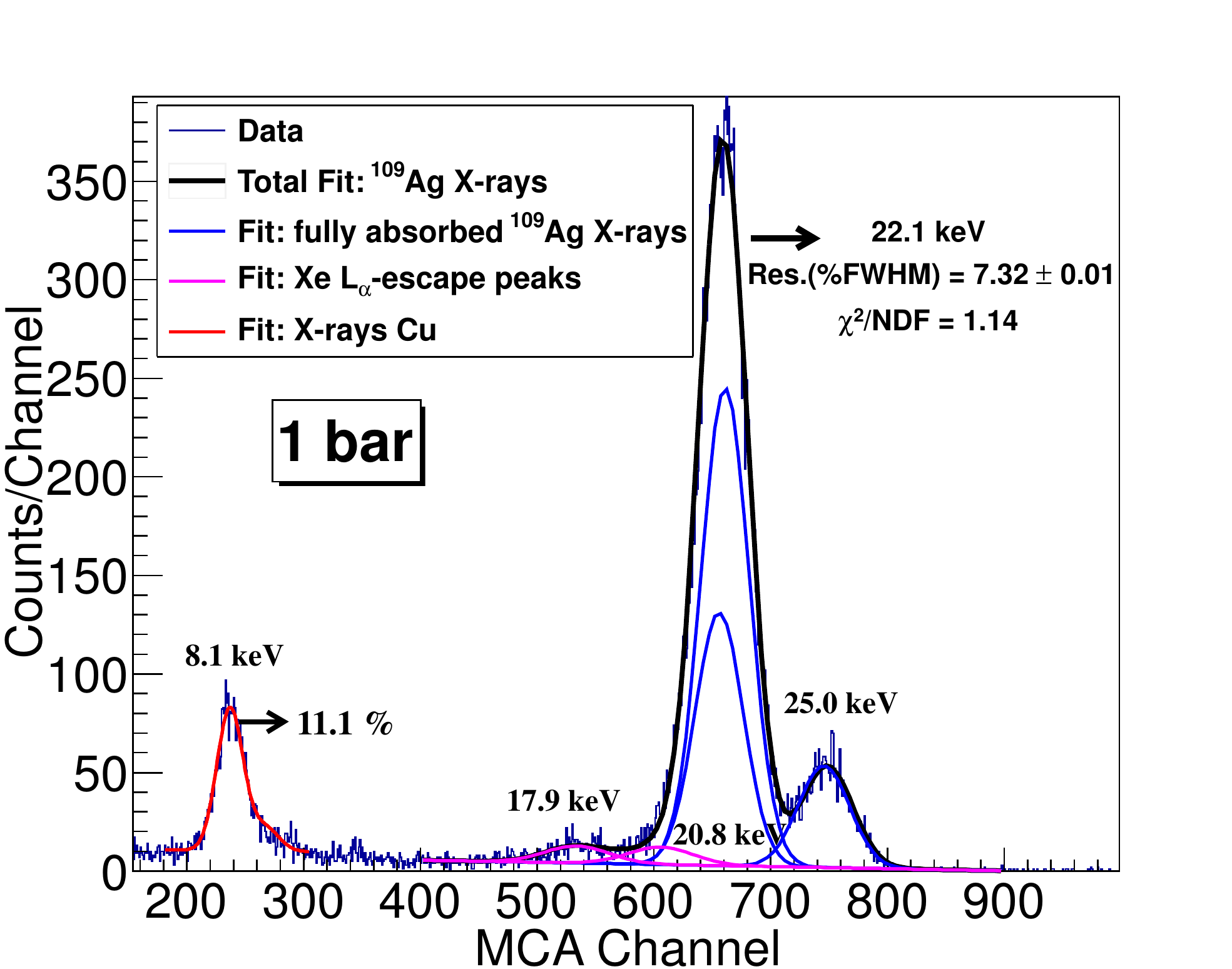}\includegraphics[scale=0.35]{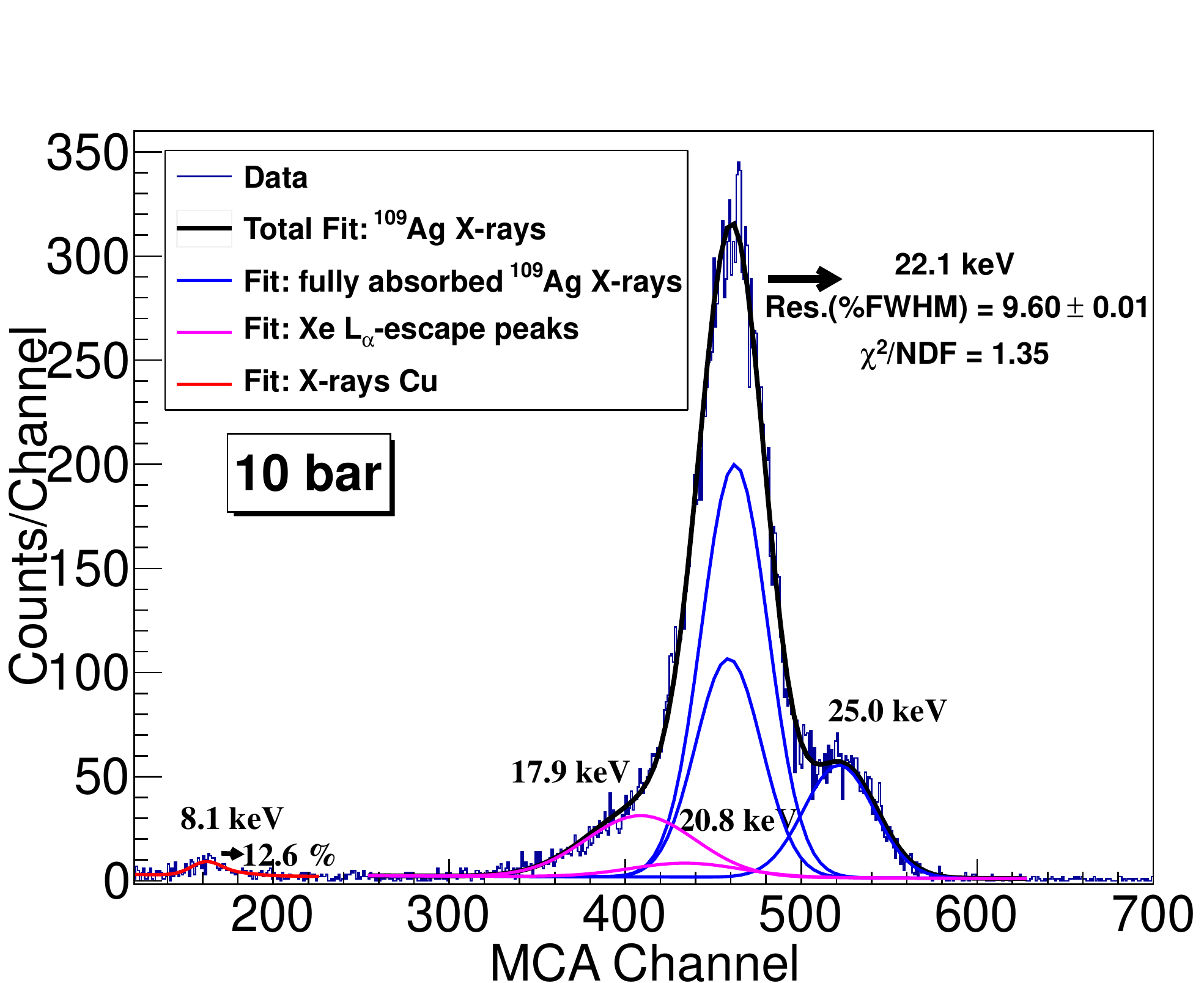}
\end{center}
\caption{\label{fig:Energy-X-rays-spectra-1-and-10bar}  X-ray energy spectra generated by a $^{109}$Cd source, in a Xe+$1.7$ \% TMA mixture at 1 bar (left) and $1.1$\% TMA at
$10$ bar (right). The fit performed to the overall $^{109}$Ag K-fluorescence lines consists in a 3-step routine. In
addition, the Cu K-fluorescence at $8.1$ keV is separately fitted to a single Gaussian function.}
\end{figure}

\subsection{Experimental procedure}\label{subsection:ExPro}

All measurements were carried out with a $^{109}$Cd collimated source placed at the center of the cathode plate.
Prior to the experimental campaign, a highly concentrated Xe+TMA mixture (93/7) was prepared in the sample cylinder.
For all subsequent measurements the mixture was constantly passed through a purifier, and its exact content adjusted by injection of fresh gas whenever needed.
Contrary to the specifications of the provider, it was observed that the SAES filter absorbs or expels TMA, 
in direct relation to the history of the previous concentrations employed. 
Hence it was necessary to wait approximately $30$ minutes until the mixture was homogenized.
Once the gas mixture was stabilized, the operating point was established.
For gas gain measurements the amplification field was systematically increased until two consecutive sparks were observed to occur within a short time ($30$ s).
At the end of each set of measurements, the TMA concentration was determined with the mass spectrometer.
The chamber and the gas system were cleaned by recovering the gas and then the complete system was pumped.

In the off-line analysis, the  $^{109}$Ag X-ray peak at $22.1\,$keV was used to obtain the
gain and the energy resolution. In Fig. \ref{fig:Energy-X-rays-spectra-1-and-10bar}, the typical energy spectrum acquired with
a $^{109}$Cd source is shown at 1 bar in a Xe+$1.7\%$\,TMA mixture
(left) and at 10 bar with $1.1$\%\,TMA (right).
In both spectra the K$_{\alpha}$ and K$_{\beta}$ lines from the Ag fluorescence are clearly distinguished.
The corresponding escape peaks from Xe are observed below the K$_{\alpha}$ line,
located at $17.9$ keV and $20.8$ keV. The Ag K$_{\alpha}$, K$_{\beta}$ lines (in blue) and the corresponding Xe escape peaks from the L-shell (in magenta) were
fitted in the energy range between $14$ and $30$\,keV using a $3$-step routine (see Fig. \ref{fig:Energy-X-rays-spectra-1-and-10bar}).
On the other hand, it is also observed the Cu K-fluorescence at $8.1$ keV, which is produced from the interaction of X-rays with the electrodes of the Micromegas.

\begin{figure}[ht]
\begin{center}
\includegraphics[scale=0.4]{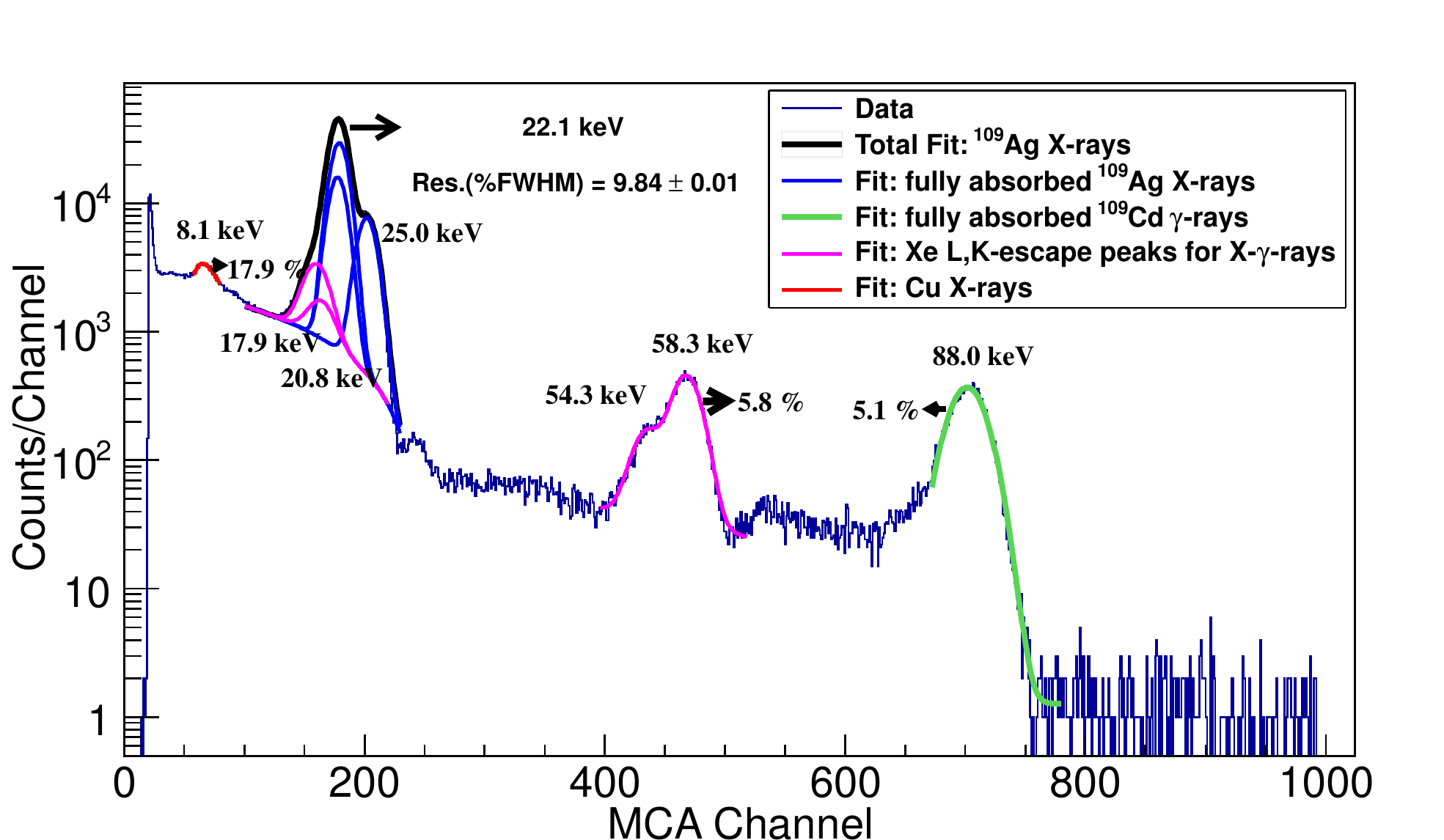}\includegraphics[scale=0.35]{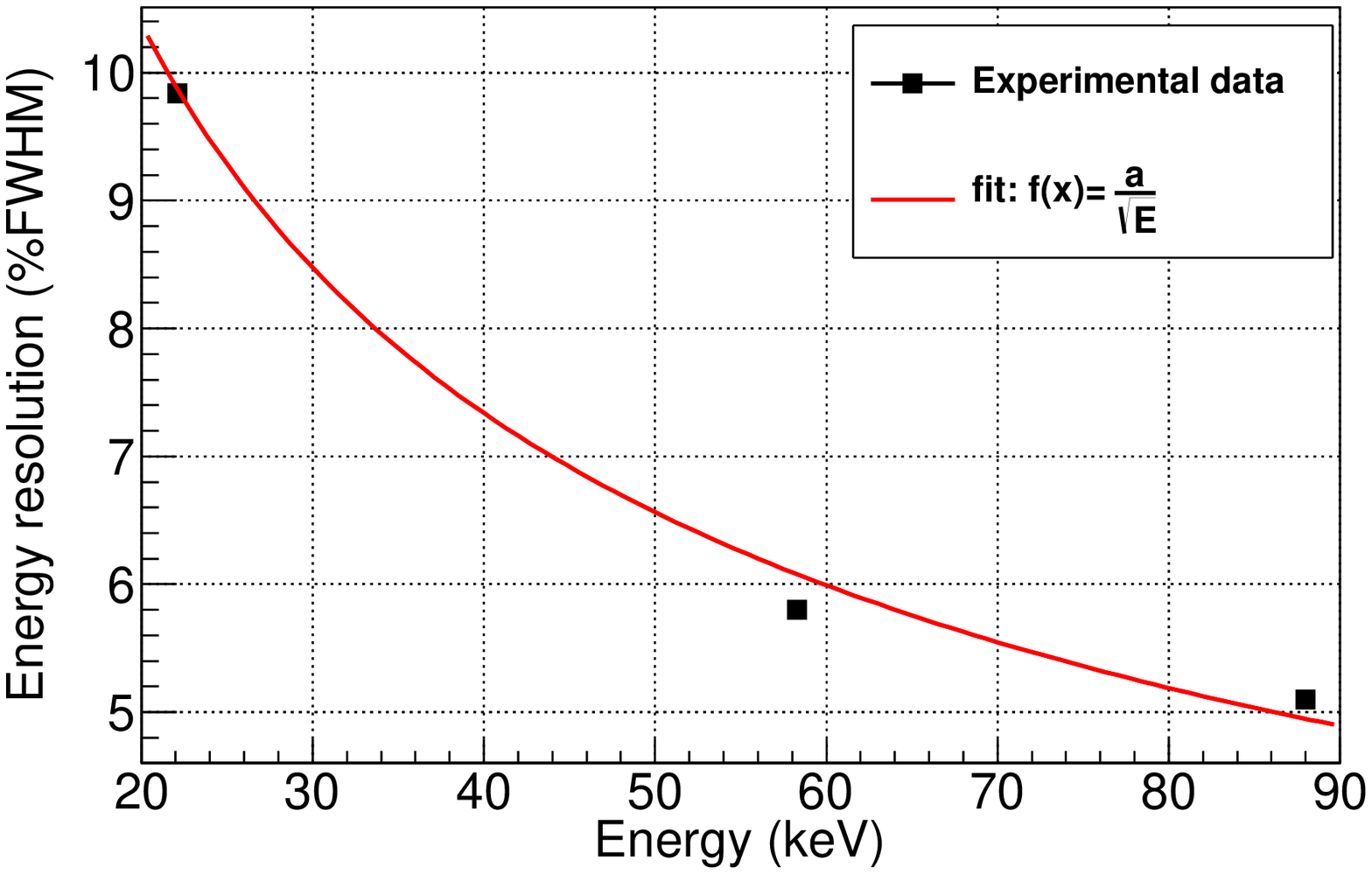}
\end{center}
\caption{Left: energy spectrum from a $^{109}$Cd source acquired at 8 bar in a Xe+$1.4$\% TMA mixture,
with $E/p=~245$ V/cm/bar in the drift region. The K-fluorescence emission and the $\gamma$-rays from $^{109}$Ag with their corresponding escape peaks of Xe are shown.
Right: behaviour of the energy resolution for various photon energies stemming from a $^{109}$Cd source, showing an approximate $1/\sqrt{E}$ scaling.
\label{fig:Energy-spectrum-at 8bar}}
\end{figure}

Fig. \ref{fig:Energy-spectrum-at 8bar} shows an energy spectrum acquired at $8$ bar using a mixture of Xe+$1.4\%$ TMA,
with a larger energy range than the previous figure. It is interesting to observe the $\gamma$-rays
from the $^{109}$Cd source at $88.04$ keV (green) and the two associated Xe escape peaks in magenta (with `escaping' energy totalling that of the K$_{\alpha}$ and K$_{\beta}$ Xe-lines).
As expected, the energy resolution is roughly inversely proportional to $\sqrt{E}$ such as shown in Fig. \ref{fig:Energy-spectrum-at 8bar} (right).

\subsection{Optimum concentration of TMA}\label{subsection:varTMA}

\begin{figure}[t]
\begin{center}
\includegraphics[scale=0.36]{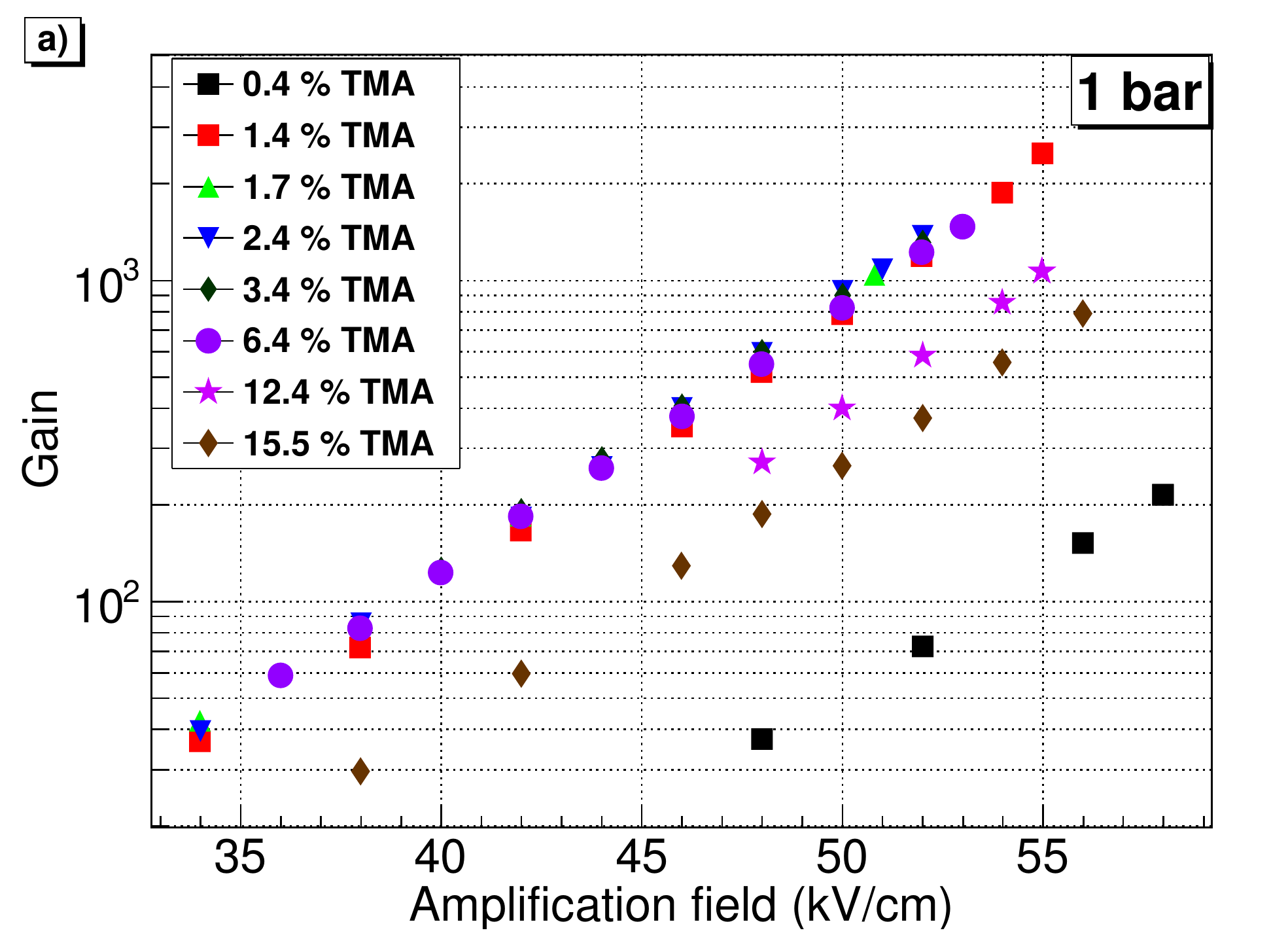}\includegraphics[scale=0.36]{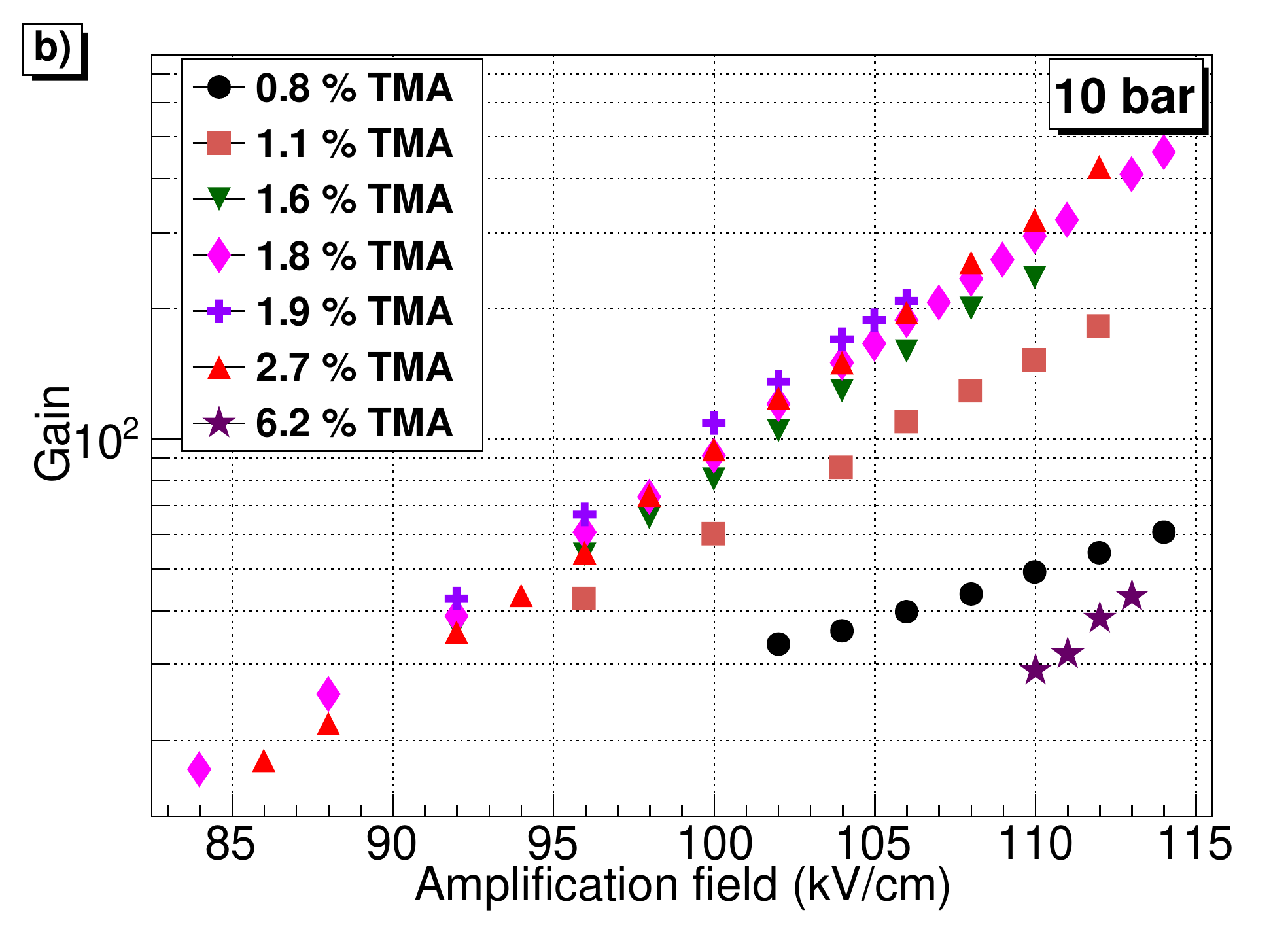}
\end{center}
\caption{Dependency of the gas gain on amplification field for different
TMA concentrations at 1 (a) and  10 (b) bar. \label{fig:GGainvsAmFi_Allpressures}}
\end{figure}

The variation in TMA concentration was performed
in different ranges: at 1 bar ($0.4\,\%-15.5\,\%$), 2 bar ($0.4\,\%-6.0\,\%$),
8 bar ($0.3\,\%-5.0\,\%$), and 10 bar ($0.8\,\%-6.2\,\%$).
The dependence of the gain with the amplification field is shown in Fig. \ref{fig:GGainvsAmFi_Allpressures} for two reference pressures:
$1$ (a) and $10$ (b) bar.
The gain curves show a linear behaviour with the amplification field in the semi-log plot, a fact often interpreted as indicative of negligible feedback. Considering the plot at 1 bar (see Fig. \ref{fig:GGainvsAmFi_Allpressures}a),
we observe that lower amplification fields must be applied
when the TMA percentage is increased from $0.4\%$ to $1.4$\%.
The curves within a concentration range from $1.4\%$ and $6.4\%$ TMA seem to overlap, suggesting
that transfer mechanisms are already fully active while the avalanche dynamics remains largely unaffected.
The tendency changes above $6.4\%$ TMA so that higher fields must be applied to obtain the same gas gain,
as also observed in neon-based mixtures \cite{PacoSaclay}.
A similar behaviour is seen at high pressures, exemplified here in the $10$ bar systematics (Fig. \ref{fig:GGainvsAmFi_Allpressures}b). The increased energy loss by inelastic collisions to TMA
molecules is the most likely explanation for this change in tendency, therefore shattering
the extra ionization obtained by Penning transfer.

\begin{figure}[ht]
\begin{centering}
\includegraphics[scale=0.35]{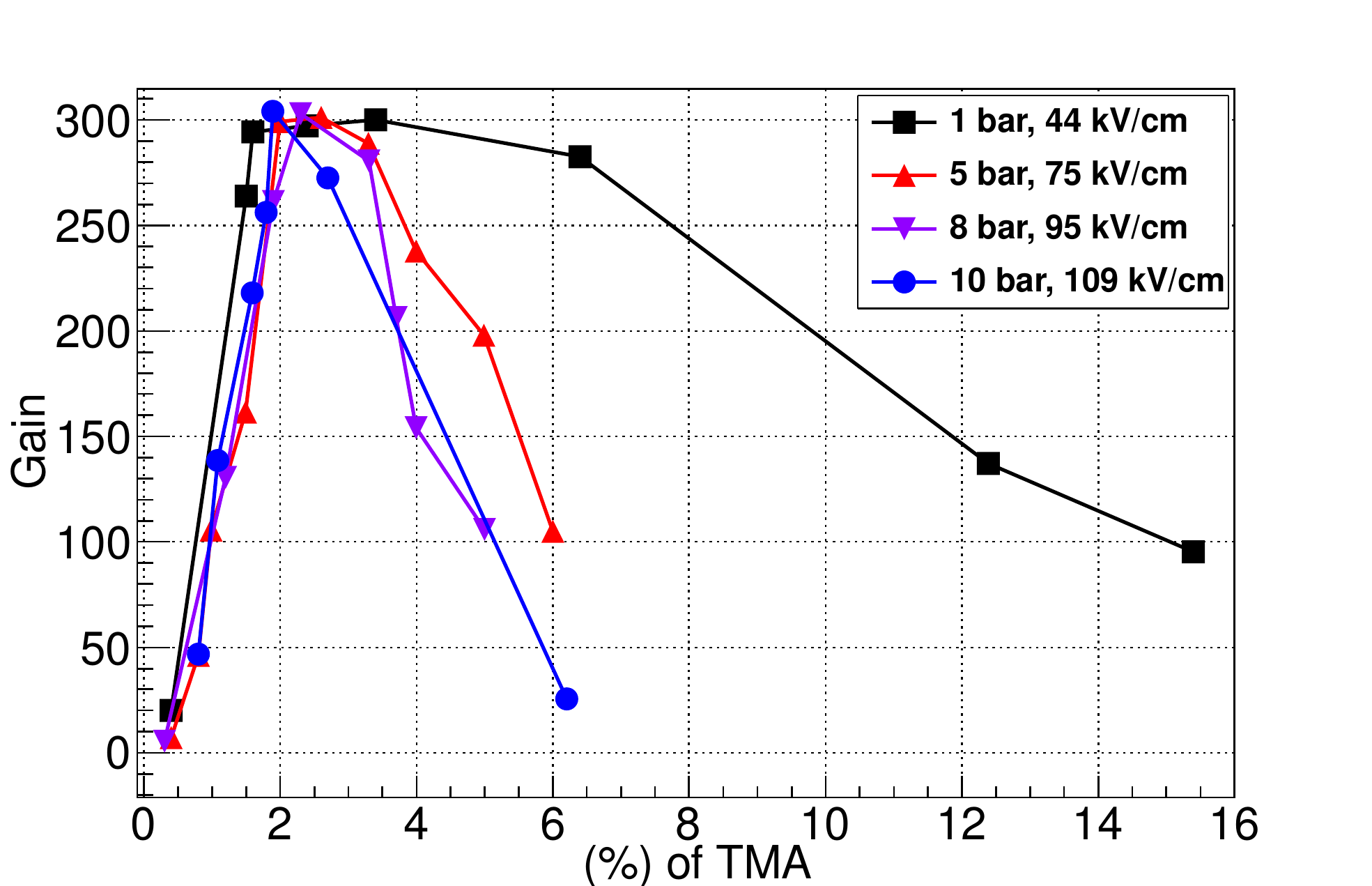}\includegraphics[scale=0.35]{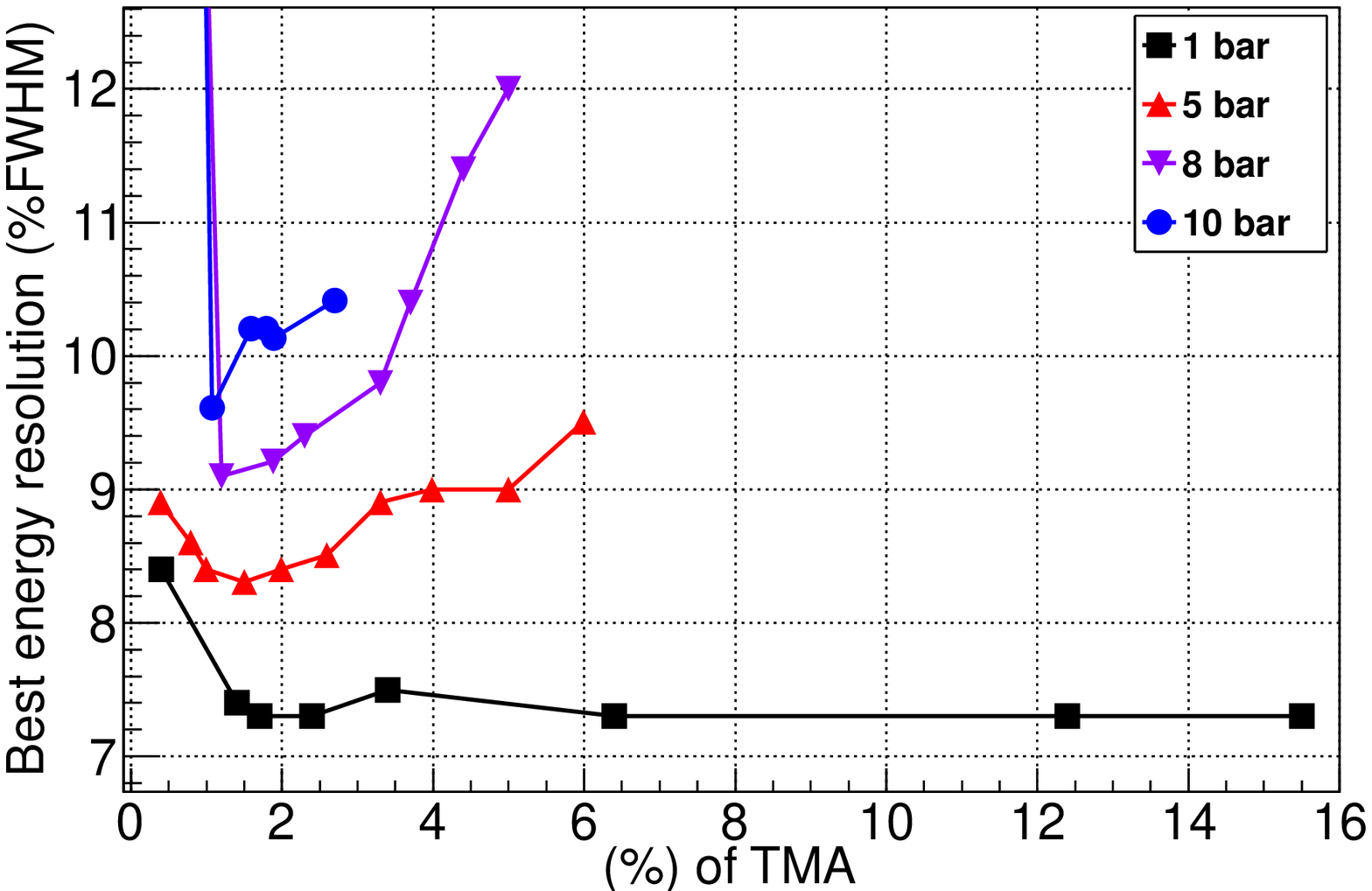}
\par\end{centering}
\caption{
Gas gain  at a fixed amplification field (left) and best energy resolution for each mixture
as a function of TMA concentration. For each pressure, an optimum region exists where the highest values in gas gain and
best energy resolutions are simultaneously achieved.
 \label{fig:AmFivsPTMA_1_5_8}}
\end{figure}

In order to do a better study of the optimum TMA concentration,
we performed linear fits of the $\ln G$ versus amplification field data for a proper interpolation or
extrapolation of the gain. The variation of the gas gain (at a constant amplification field) with the percentage of TMA is
shown in Fig. \ref{fig:AmFivsPTMA_1_5_8} (left). At each pressure the gas gain rapidly rises when small quantities of TMA are added up to
values around $2\%$ TMA. Further, a range where the gain remains roughly constant occurs between $2\%$ and $3\%$ TMA, except for $1$ bar
where the gain remains constant in a larger range ($2\%$-$6\%$). The gas gain decreases sharply thereafter.
At low pressures up to atmospheric, the rapid rise in gas gain at constant amplification field has also been observed using Xe+$2$, $3$ dimethyl-2-butene as Penning additive, showing similar dependencies with the additive at $1$ bar \cite{DMB}.
This great increase is a strong evidence that Penning effect takes place.

Figures showing the energy resolution as a function of amplification field are not presented in this work (but can be
found in \cite{Diana}).
The best energy resolution for each mixture is shown, however, in Fig. \ref{fig:AmFivsPTMA_1_5_8} (right)
as a function of the TMA concentration. In general, we can see that the energy resolution improves with the TMA concentration, the best values of energy resolution being found between $1\%$ and $2.5\%$ for each pressure.
Therefore, we can conclude that the optimal TMA percentage ranges from $1.5\%$ to $2.5\%$ where the best energy resolutions and the highest gains (at constant field) are obtained.

\subsection{Varying the pressure}\label{subsection:VarPre}

\begin{figure}[ht]
\begin{centering}
\includegraphics[scale=0.36]{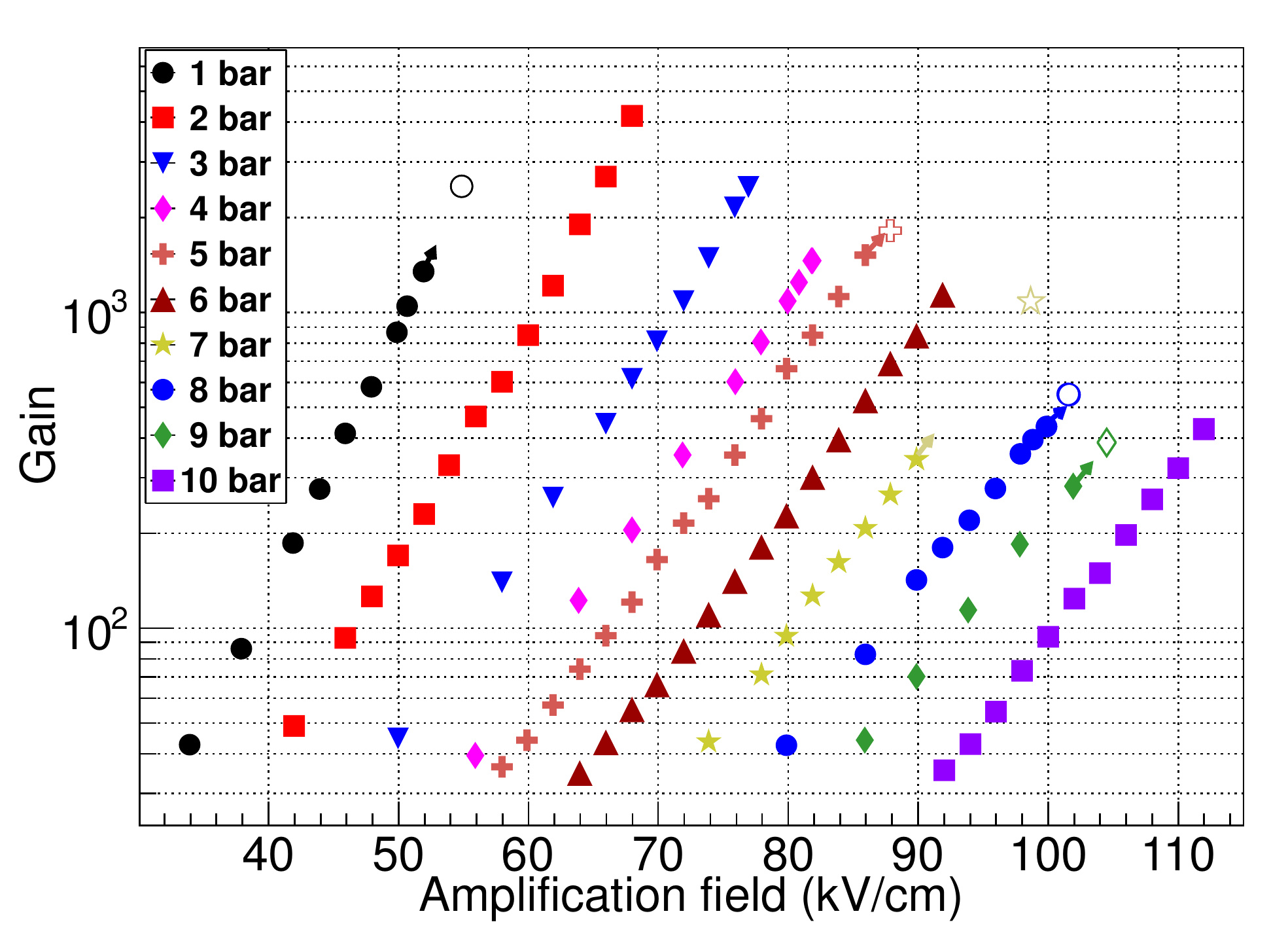}\includegraphics[scale=0.38]{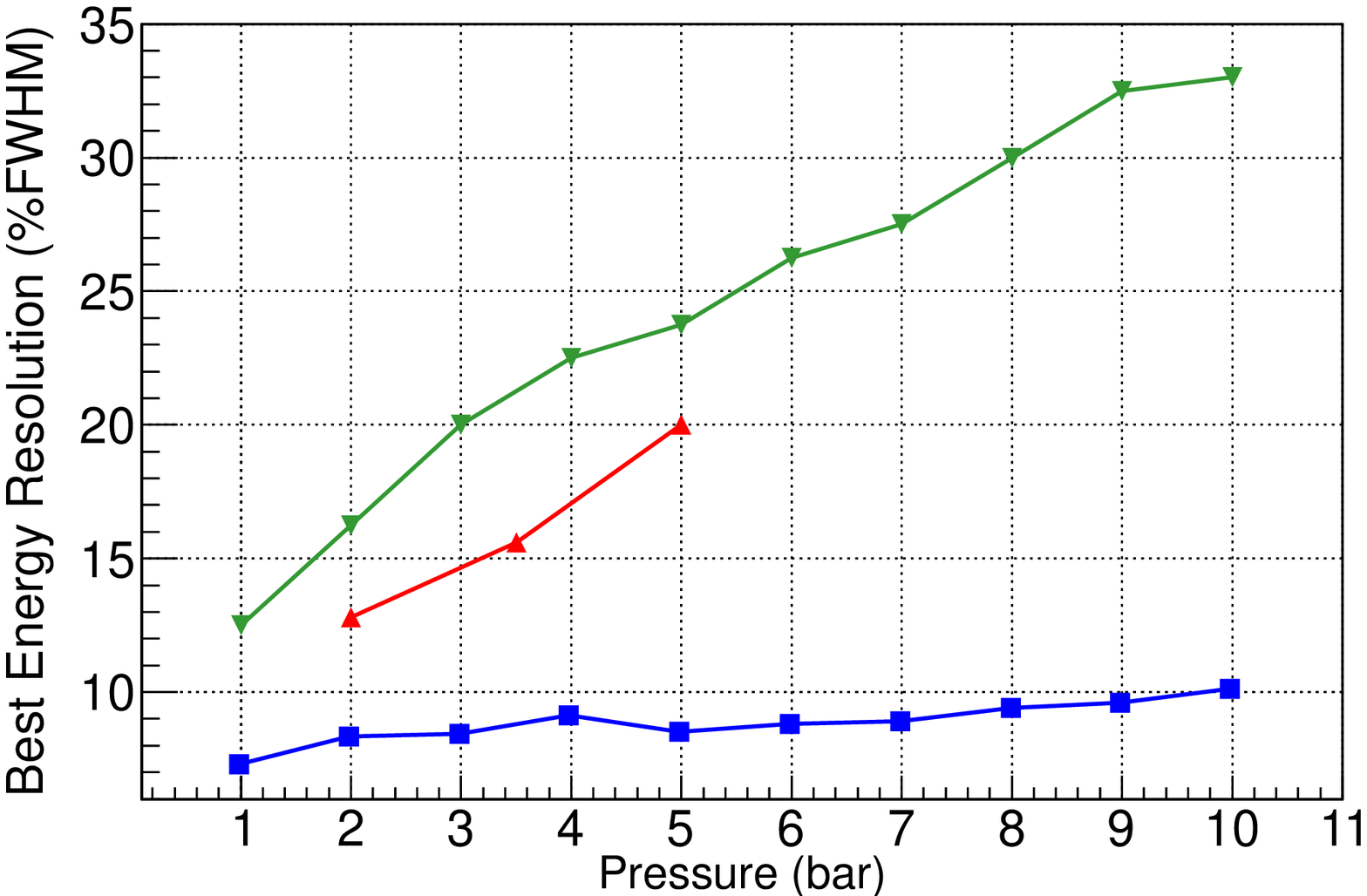}
\par\end{centering}
\caption{Left: Dependence of the gain with the amplification field. The empty marker represents the maximum gain reached for each pressure, in case it was not determined for the given conditions but for a slightly different gas mixture around the optimum. Right: Dependence of the best energy resolution with pressure at 22.1 keV, for data from \cite{Coimbra} (\textcolor{green}{$\blacktriangledown$}),
from \cite{XeZaragoza} (\textcolor{red}{$\blacktriangle$}) and from this work (\textcolor{blue}{$\blacksquare$}).
\label{fig:GGainvsAmFi}}
\end{figure}

In this section we present the results of the gas gain and the energy resolution
obtained when the pressure is varied from $1$ to $10$ bar,
using TMA concentrations within the optimal range that was estimated in the previous section (from $1.5\,\%$ to $2\,\%$ TMA).
Gas gain curves are shown in Fig. \ref{fig:GGainvsAmFi} (left).
It is observed that the maximum gain drops nearly exponentially for pressures above $2$ bar,
down to $\sim400$ at $10$ bar. However, the maximum gain at any pressure is still at least
a factor 3 higher than for Micromegas operated in pure Xe \cite{Coimbra}.

The results of this work are compared in Fig. \ref{fig:GGainvsAmFi} (right)
with previous measurements with Micromegas detectors in pure Xe:
with the same setup used in this study (\textcolor{red}{$\blacktriangle$}) \cite{XeZaragoza}
and with a different one (\textcolor{green}{$\blacktriangledown$}) \cite{Coimbra}.
The energy resolution achieved at $22.1$ keV is substantially better in this work,
going down to $7.3\,\%$ ($9.6\,\%$) FWHM at $1$ ($10$) bar.
This fact translates into an improvement of a factor $2$ ($3$) at $1$ ($10$) bar as compared to previous
measurements in pure Xe. We can infer that the addition of TMA to Xe reduces the avalanche fluctuations
both due to the efficient Penning transfer between Xe excited states and TMA molecules (causing ionization
of the latter), and to the efficient suppression of photon feedback.

\section{Coincidence setup}\label{section:NEXT-0-MMsilicon}

\begin{figure}[hb]
\begin{centering}
\includegraphics[scale=0.32]{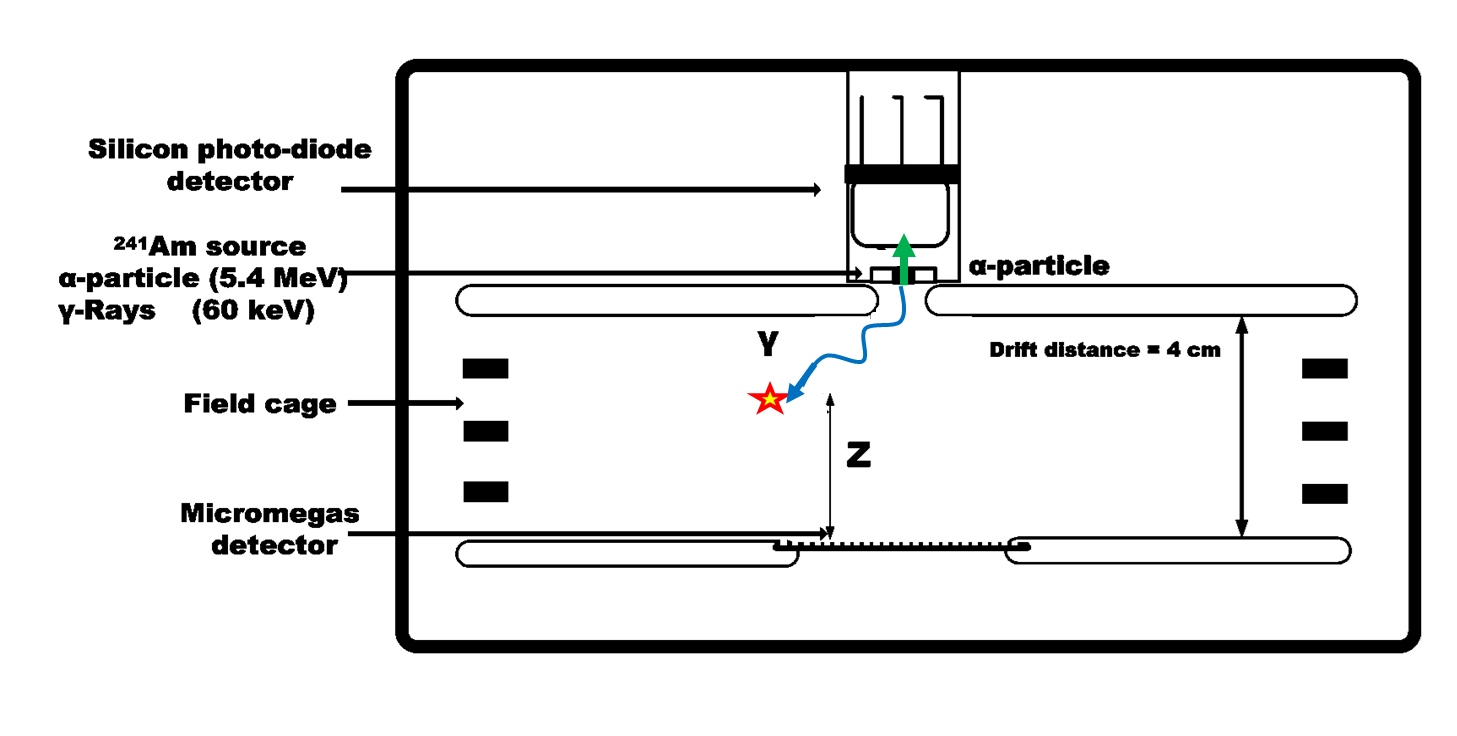}\includegraphics[scale=0.37]{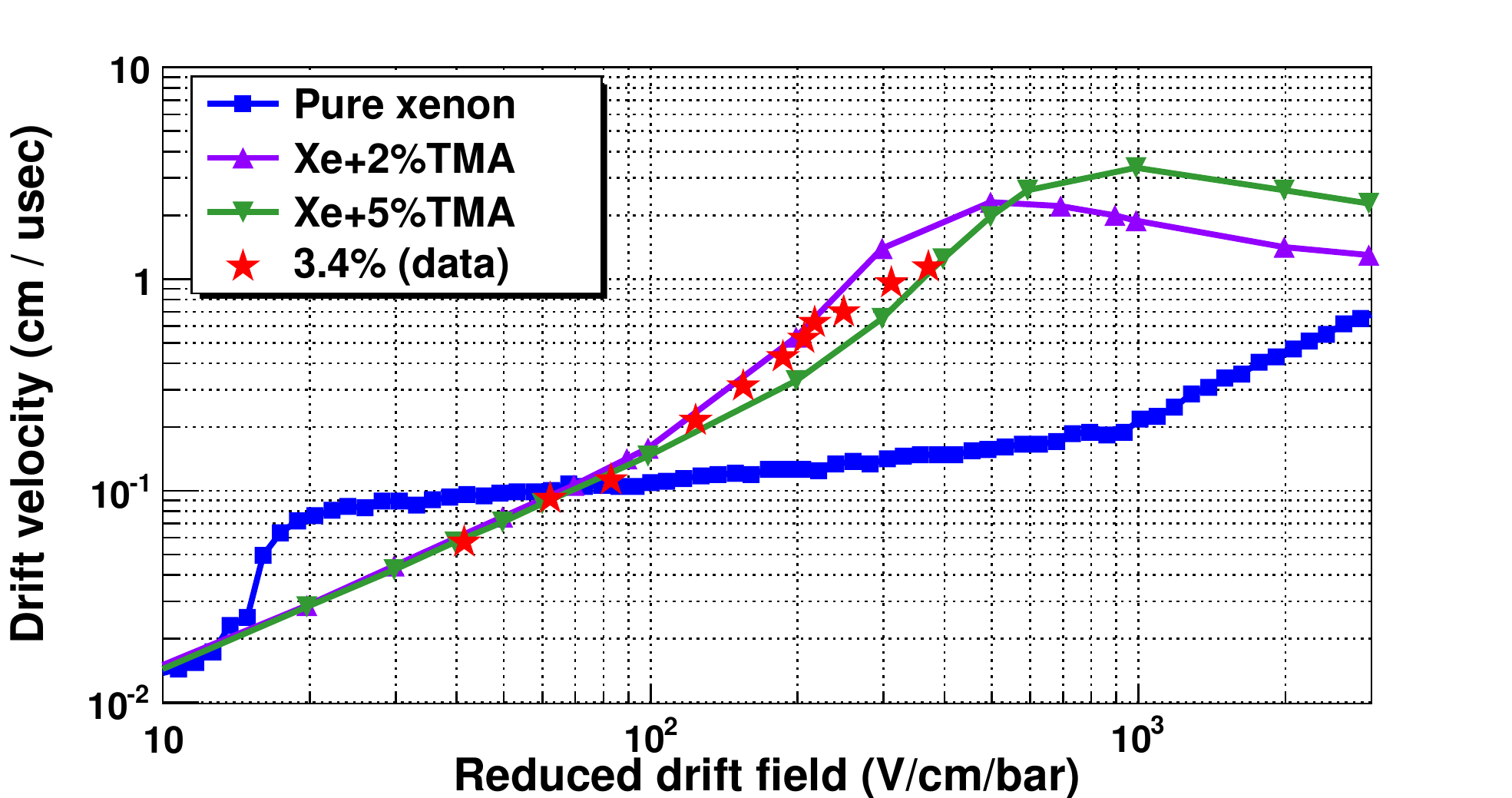}
\par\end{centering}
\caption{Left: modified setup designed for measuring the drift velocity and attachment.
Right: measured drift velocity together with the ones calculated with Magboltz-9 for Xe+TMA
mixtures and for pure Xenon. The red markers represent the drift velocity measured for a Xe+$3.4\%$ TMA mixture.
On the other hand, the violet and the green curve represent the drift velocity calculated with Magboltz for Xe+$2\%$ and Xe+$3\%$ TMA mixtures, respectively.
\label{fig:SetUpSi}}
\end{figure}

The small-TPC setup was modified to measure the drift velocity and attachment effects in Xe+TMA mixtures.
A schematic view of the setup used for these measurements is shown in Fig. \ref{fig:SetUpSi} (left).
A silicon photo-diode detector together with an $^{241}$Am source was encapsulated into a plastic piece (made of POM, with low outgassing), and then installed inside the TPC.
The $^{241}$Am source emits in coincidence an $\alpha$-particle and a $\gamma$-photon which are detected by the silicon diode and the Micromegas detectors, respectively.
Both signals are pre-amplified and registered by a Tektronix oscilloscope, that allows to select the coincidences.
With this configuration we obtain a coincidence system which provides the $t_0$ of each event, hence the drift time and the drift velocity can be obtained.
The drift time of each event is calculated from the temporal difference between the $\alpha$ and $\gamma$ signals.
The drift velocity is then obtained from the overall range of drift times ($\Delta{T}$), that spans the full drift region, and the (fixed) drift distance ($\Delta{x}$) as $v_e = \Delta{x}/\Delta{T}$. We have in this way determined the drift velocity for various Xe+TMA mixtures and pressures. In Fig. \ref{fig:SetUpSi} (right) our preliminary results and a comparison with Magboltz calculations are compiled.


On the other hand, the dependence of Micromegas amplitude with the temporal distance between signals can be used to study attachment effects in our gas.
In principle, in absence of attachment the signal amplitude should be independent from the drift time, but will show an exponential behaviour otherwise, whose exponent is usually referred to as the inverse of the electron lifetime, $\tau_e$. We have preliminary estimated an electron lifetime (1/(attachment coefficient $\times$ drift velocity)) larger than $\tau_e=4$ ms with this setup (up to 6 bar). Final results will be present elsewhere.

\section{Conclusions and outlook}\label{section:Conclusions}

We have performed systematic measurements of gain and energy resolutions in a small TPC filled with Xe-TMA on a large
range of concentrations and pressures.
We have found that, between 1 and 10 bar, the Penning transfer is optimal for Micromegas operation in the $1.5\%$-$2.5\%$ TMA concentration range, allowing to obtain the best energy resolutions and the highest gain increase at constant field.
Energy resolutions down to $7.3\%$ ($9.6\%$) FWHM at $1$ ($10$) bar for  $22.1$ keV can be achieved, which imply
an improvement of a factor $2$ ($3$) with respect to values previously obtained also with microbulk technology, but in pure Xe \cite{Coimbra}.
This result extrapolates into an energy resolution of $0.7\%$ ($0.9\%$) FWHM at the $Q_{\beta\beta}$ value of Xe for $1$ ($10$) bar,
and therefore opens very good prospects for 0${\nu\beta\beta}$ decay experiments.
In addition, we have performed a modification of this setup to measure the drift velocity and the attachment effects.
First measurements of the drift velocity in Xe+TMA mixtures have been made and show good agreement with Magboltz calculations.

As part our current work, we have performed measurements in Xe-TMA mixtures in a medium TPC (80 l) at 1 bar using microbulk technology. 
Preliminary results have permitted to observe the first tracks reconstructed of X-rays, $\gamma$-emission and background events.
The detector performance and the purification system are currently being studied for future improvements. 
A paper is under preparation to describe in detail the setup, a complete study of the topology of the tracks, energy resolution and gas gain \cite{NEXT-2013}.

\section*{Acknowledgments}
We are grateful to our colleagues of the groups of the University of
Zaragoza, CEA/Saclay as well as the NEXT and RD-51 collaborations.
We thank R. de Oliveira and his team at CERN for the manufacturing of the
microbulk readouts. We acknowledge support from the European
Commission under the European Research Council T-REX Starting Grant
ref. ERC-2009-StG-240054 of the IDEAS program of the 7th EU Framework
Program. We also acknowledge support from the Spanish Ministry of Economics and Competitiveness (MINECO),
under contracts ref. FPA2008-03456 and FPA2011-24058, as
well as under the CUP project ref. CSD2008-00037 and the CPAN project
ref. CSD2007- 00042 from the Consolider-Ingenio 2010 program of the
MICINN. Part of these grants are funded by the European Regional
Development Fund (ERDF/FEDER). F.I. acknowledges the support from the Eurotalents program.

\end{document}